\documentclass[3p,times,twocolumn]{elsarticle}

\usepackage{amssymb}
\usepackage[figuresright]{rotating}

\begin{document}

\begin{frontmatter}




%%--- EDIT Title ---
\title{Flavor Oscillations in Core-Collapse Supernovae}
%%-------------------

%%--- EDIT Author information ---
%%     Author information
%% \author[label1,label2]{<author name>}ollective
\author[label1]{A.B.~Balantekin}
\address[label1]{Physics Department, University of Wisconsin, Madison WI 53706 USA}
%%---------------------------------

%%--- EDIT Abstract ---
\begin{abstract}
Core collapse supernovae are unique laboratories to study many aspects of neutrino physics. 
The vicinity of the proto-neutron star in a core-collapse supernova is characterized by large  
matter and neutrino densities. A salient feature of this region is the impact of neutrino-neutrino interactions.  Properties of the ensuing non-linear many-neutrino system are examined with a particular emphasis on its collective behavior and its symmetries. The impact of neutrino  properties and interactions on the r-process nucleosynthesis that may take place in the supernova environment is discussed. 
\end{abstract}
%%------------------------

%%--- EDIT Keywords ---
\begin{keyword}
Supernova neutrinos \sep collective neutrino oscillations
\end{keyword}
\end{frontmatter}

% \linenumbers

%%--- EDIT Main text ---
\section{Introduction}
\label{sec:introduction}
\begin{figure}
\begin{center}
	\resizebox{\linewidth}{!}{\includegraphics{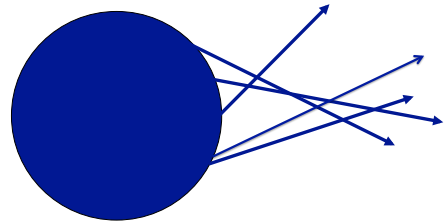}}
	\caption{Schematic illustration of the neutrino-neutrino interactions near the neutrinosphere.}
	\label{fig:1}
\end{center}
\end{figure}

Core collapse supernovae are unique laboratories to study many aspects of neutrino physics. Following the collapse, almost the entire gravitational binding energy of the progenitor star is emitted in neutrinos. The collapse results in very large values of Fermi energy for electrons and electron neutrinos, about $10^{57}$ units of electron lepton number. Neutrinos then transport entropy and electron lepton  number\footnote{Note that since $\mu$ and $\tau$ neutrinos are pair produced they carry no net muon or tau lepton number.} away from the proto-neutron star. Neutrinos 
dominate the energetics of core-collapse supernovae; they carry about 10\% of the progenitor star's rest mass, $10^{53}$ ergs, in contrast to 
the total optical and kinetic energy which about only one percent of this amount. Since the diffusion time of the neutrinos (slightly less than ten 
seconds) is much longer than typical time scale of weak interactions, the sheer number of neutrinos present in the proximity of the proto-neutron star enable many-body aspects of neutrino physics emerge. Resulting collective neutrino oscillations allow testing a sector of the Standard Model that cannot be tested elsewhere, namely the weak interaction between two neutrinos, as schematically depicted in Figure \ref{fig:1}. There are several excellent review articles which may serve as starting points on the expansive literature exploring physics opportunities with supernova neutrinos 
\cite{Duan:2009cd,Raffelt:2010zza,Duan:2010bg}. 

In a supernova, matter-enhanced neutrinos oscillations (the MSW effect) is operational not only for the neutrinos, but, under certain conditions, for antineutrinos as well. 
However, collective effects (especially non-diagonal neutrino-neutrino interaction terms) dominate the neutrino transport much deeper than the MSW effect.   

Core-collapse supernovae are also plausible sites for the r-process nucleosynthesis. The parameter controlling nucleosynthesis, electron fraction or equivalently neutron-to-proton ratio, is determined by the neutrino capture rates. Interactions of the neutrinos and antineutrinos streaming out of the core both with nucleons and seed nuclei determine the neutron-to-proton ratio. Hence to understand core-collapse supernovae and the r-process nucleosynthesis it may host it is crucial to fully understand neutrino properties and interactions \cite{Balantekin:2003ip}. 

If there are sufficiently energetic electron neutrinos present among the neutrinos emitted from the cooling proto-neutron star, they can convert neutrons into protons. Those neutrons could otherwise initiate the r-process nucleosynthesis. Instead they get bound in helium nuclei, an energetically favorable situation, dropping out of the r-process. This "alpha effect" can cause significant increases in the electron fraction, halting the formation of r-process elements \cite{McLaughlin:1997qi}. 

\section{The Mikheyev, Smirnov, Wolfenstein (MSW) effect in a core-collapse supernova}

Using the following parameterization of the neutrino mixing matrix: 
\begin{eqnarray}
\label{aaa1}
 &&\hspace{-1cm}{\bf T} = {\bf T}_{23}{\bf T}_{13}{\bf T}_{12} \nonumber \\
&&\hspace{-1cm}=  \left(
\begin{array}{ccc}
 1 & 0  & 0  \\
  0 & C_{23}   & S_{23}  \\
 0 & -S_{23}  & C_{23}  
\end{array}
\right)
\left(
\begin{array}{ccc}
 C_{13} & 0  & S_{13} e^{-i\delta_{CP}}  \\
 0 & 1  & 0  \\
 - S_{13} e^{i \delta_{CP}} & 0  & C_{13}  
\end{array}
\right) \nonumber \\
&& \times
\left(
\begin{array}{ccc}
 C_{12} & S_{12}  & 0  \\
 - S_{12} & C_{12}  & 0  \\
0  & 0  & 1  
\end{array}
\right)
\end{eqnarray}
where $C_{ij} = \cos \theta_{ij}$, $S_{ij} = \sin \theta_{ij}$, and $\delta_{CP}$ is the CP-violating phase, the MSW evolution equations can be written as 
\begin{equation} 
\label{CProt}
i \frac{\partial}{\partial t} 
\left(
\begin{array}{c}
  \Psi_e \\
 \Psi_{\mu} \\
  \Psi_{\tau} 
\end{array}
\right) 
= {\bf H} 
\left(
\begin{array}{c}
  \Psi_e \\
  \Psi_{\mu} \\
  \Psi_{\tau} 
\end{array}
\right) 
\end{equation}
with
\begin{equation}
\label{hmsw}
{\bf H} = 
{\bf T}
\left(
\begin{array}{ccc}
E_1  & 0  & 0  \\
0  & E_2  & 0  \\
0  &  0 & E_3  
\end{array}
\right) {\bf T}^{\dagger}  + 
\left(   
\begin{array}{ccc}
 V_{e \mu} & 0  & 0  \\
 0 & 0 & 0  \\
0  & 0   &  V_{\tau \mu}  
\end{array}
\right) .
\end{equation}
In Eq. (\ref{hmsw}), by dropping a term proportional to the identity, $V_{\mu \mu}$ is chosen to be zero.  The non-zero potentials in this equation are the widely used tree-level contribution \cite{Wolfenstein:1977ue}
\begin{equation}
  \label{wolfen1}
V_{\mu e} (x) = \sqrt{2} G_F  N_e (x) ,
\end{equation}
where $N_e$ is the effective electron density, i.e. the difference between electron and positron densities, 
and the Standard Model loop correction \cite{Botella:1986wy}, 
\begin{eqnarray}
\label{loop}
V_{\tau \mu} &=& -  \frac{3 \sqrt{2} G_F\alpha}{\pi \sin^2 \theta_W}  \left( \frac{m_{\tau}}{m_W}\right)^2 \nonumber \\
&\times&
\left\{ (N_p + N_n) \log \frac{m_{\tau}}{m_W} + \left( \frac{N_p}{2} + \frac{N_n}{3}  \right) \right\} .
\end{eqnarray}
Especially when the muon and tau neutrino fluxes emitted from the proto-neutron star differ, loop-correction contributions to the neutrino potential may play an important role \cite{Akhmedov:2002zj}. 

Here we assumed that neutrinos interact with uniform, monotonically-varying background matter, ignoring any background fluctuations. In the presence of fluctuations, \begin{equation}
\label{fluc}
N_e = < N_e> + {\rm \> fluctuating \> part} ,
\end{equation}
where $<N_e>$ is the fluctuation averaged part of the electron density, one can observe interesting effects \cite{Loreti:1994ry}.
Effects of such random density fluctuations on two-neutrino flavor transformations in the post-core-bounce supernova environment was first examined long ago \cite{Loreti:1995ae}. Recently there has been renewed interest in exploring the effects of turbulence and density fluctuations in core-collapse supernovae \cite{Cherry:2011fm}. 

With three flavors, one has to explicitly include the CP-violating phase in discussing supernova neutrinos. However, it is rather straightforward to show that the CP-violating phase factorizes out in the neutrino evolution Hamiltonian if only the tree level neutrino potential of Eq. 
(\ref{wolfen1}) is used. 
This factorization gives us interesting sum rules:
 Electron neutrino survival probability, $P (\nu_e \rightarrow \nu_e)$ is independent of the value of the CP-violating phase, $\delta$; or equivalently, in the absence of sterile neutrino mixing, 
the combination $P (\nu_{\mu} \rightarrow \nu_e) + P (\nu_{\tau} \rightarrow \nu_e)$ at a fixed energy is independent of the value of the CP-violating phase \cite{Balantekin:2007es}.  It is possible to derive similar sum rules for other amplitudes \cite{Kneller:2009vd}. Discussions of the breakdown of this formula when sterile neutrinos are present is given in Ref. \cite{Palazzo:2011rj} and the impact of the CP-violating phase on collective effects in Ref. \cite{Gava:2008rp}

The discussion above only pertains to three active flavors, ignoring the possibility of mixing between active and sterile components in the neutrino sector. 
Recently, an analysis of the short-baseline reactor neutrino experiments suggested the possibility of a discrepancy between observations and the predicted antineutrino flux \cite{Mention:2011rk}, implying contributions from sterile neutrinos. This interpretation is consistent with anomalous results from various other neutrino experiments and inferences from cosmology \cite{Abazajian:2012ys}. It is worth pointing out that the reactor neutrino flux deficiency could also be a consequence of not yet accounted for systematic errors in the reactor flux calculations. It was suggested that such a deficit can be explained with a single additional sterile neutrino state with mass splitting of the order of  $\sim 1 {\rm eV}^2$ and a rather small mixing with active states \cite{Giunti:2011cp}. Incidentally, a sterile neutrino mixing with the electron neutrino with similar parameters would reduce the electron neutrino flux in a supernova eliminating the alpha effect and enabling the r-process nucleosynthesis to proceed \cite{McLaughlin:1999pd}. Impact of eV-mass sterile neutrinos on neutrino-driven supernova outflows was recently studied in 
Ref. \cite{Tamborra:2011is}. Such sterile states may help supernova explosions \cite{Hidaka:2007se}.  

Finally, it is worth pointing out that, in addiction to the direct observation of the neutrinos immediately following a supernova explosion, it should also be possible to observe the diffuse background of the integrated neutrino flux from past supernovae \cite{Beacom:2010kk}. 
Clearly if the diffuse supernova neutrino background is not detected, then new physics will be required. 

\section{Collective neutrino oscillations} 
We first consider, for simplicity, only two flavors of neutrinos: electron neutrino, $\nu_e$, and 
another flavor, which we denote $\nu_x$ (typically a combination of the muon and tau neutrinos). Introducing the creation and annihilation operators for a neutrino 
with three momentum ${\bf p}$, we can write down the generators of the neutrino flavor isospin algebras   
\cite{Balantekin:2006tg}: 
\begin{eqnarray}
J_+({\bf p}) &=& a_x^\dagger({\bf p}) a_e({\bf p}), \> \> \>
J_-({\bf p})=a_e^\dagger({\bf p}) a_x({\bf p}), \nonumber \\
J_0({\bf p}) &=& \frac{1}{2}\left(a_x^\dagger({\bf p})a_x({\bf p})-a_e^\dagger({\bf p})a_e({\bf p})
\right). \label{su2}
\end{eqnarray}
Different algebras representing neutrinos with different three momenta commute. 
The integrals of these operators over all possible values of momenta generate a global flavor 
isospin algebra. (It is an SU(2) algebra). 
One can then show that the neutrino mixing is simply a transformation under the associated global flavor isospin group: 
\[
\hat{U}^\dagger a_1(\mathbf{p},s)\hat{U}  = a_e(\mathbf{p},s) 
\]
with 
\begin{equation}
\label{mixalg}
\hat{U} =e^{\sum_p \tan \theta J_p^+} \; e^{- \sum_p \ln(\cos^2 \theta) J_0^p} \; e^{-\sum_p \tan \theta J_p^-}  ,
\end{equation}
where numerical indices (1 and 2) refer to the creation and annihilation operators in the mass basis. 
 Using the SU(2) generators in Eq. (\ref{su2}), it can be shown that 
the Hamiltonian for a neutrino propagating through matter takes the form  
\begin{eqnarray}
\label{msw}
&&\hspace{-1cm} H_{\nu} = \nonumber \\&&\hspace{-1cm} \int d^3{\bf p} \frac{\delta m^2}{2p} \left[
\cos{2\theta} J_0({\bf p}) + \frac{1}{2} \sin{2\theta}
\left(J_+({\bf p})   + J_-({\bf p})\right) \right] \nonumber \\ &&-  \sqrt{2} G_F \int d^3{\bf p} 
\> N_e \>  J_0({\bf p}).  
\end{eqnarray}
In Eq. (\ref{msw}), the first integral represents the neutrino mixing and the second integral 
represents the neutrino forward scattering off the background matter. (For a detailed description see references \cite{Pantaleone:1992eq} through \cite{Saviano:2012yh} ). 
In writing the Eq. (\ref{msw})   
a term proportional to identity is omitted as such terms do not contribute to the phase interference in neutrino oscillations. 
Neutrino-neutrino interactions are described by the Hamiltonian 
\begin{equation}
\label{nunu}
\hspace{-0.6cm} H_{\nu \nu} = \frac{\sqrt{2}G_F}{V} \int d^3{\bf p} \> d^3{\bf q} \>  (1-\cos\vartheta_{\bf pq}) \> {\bf
J}({\bf p}) \cdot {\bf J}({\bf q}) ,
\end{equation}
where $\vartheta_{\bf pq}$ is the angle between neutrino momenta {\bf p} and {\bf q} and V 
is the normalization volume.  Note that the presence of the $(1-\cos\vartheta_{\bf pq}) $ term in 
the integral above is crucial to recover the effects of the Standard Model weak  
interaction physics in the most general situation\footnote{For a recent discussion of the impact of the physics 
beyond the Standard Model 
on matter-enhanced neutrino oscillations see \cite{Balantekin:2011ft} and references therein.}. 
In the extremely idealized case of 
isotropic neutrino distribution and a very large number of neutrinos, this term may average to a 
constant and the neutrino-neutrino interaction 
Hamiltonian simply reduces to the Casimir operator of the global SU(2) algebra. In the literature this is called the "single-angle 
approximation".
Single-angle approximation is unlikely to hold  
in core-collapse supernovae for all locations and epochs, but it is the appropriate procedure for the Early Universe 
\cite{Kostelecky:1993ys,Abazajian:2002qx}. It should be noted that even in this limit the problem of evaluation the neutrino evolution 
operator is far from trivial. 

The discussion above pertains to a gas comprised of only neutrinos without any antineutrinos. Including antineutrinos 
requires introduction of a second set of SU(2) algebras, separate ones for each neutrino and antineutrino momenta. Similarly for three flavors two 
sets of SU(3) algebras 
are needed \cite{Sawyer:2005jk}. Both extensions are straightforward, but rather tedious to implement in practice. 

Defining the auxiliary vector quantity 
\begin{equation}
\hat{B} = (\sin2\theta,0,-\cos2\theta), 
\end{equation}
the total Hamiltonian with two flavors, containing one- and two-body interaction terms, can be written in the form 
\begin{eqnarray}
\label{total}
\hat{H}_{\mbox{\tiny total}} &=& H_{\nu} + H_{\nu \nu}  \nonumber \\ 
&=& \left(
\sum_p\frac{\delta m^2}{2p}\hat{B}\cdot\mathbf{J}_p  - \sqrt{2} G_F 
N_e  \mathbf{J}_p^0  \right) \nonumber \\
&+& \frac{\sqrt{2}G_{F}}{V}\sum_{\mathbf{p},\mathbf{q}}\left(1- 
\cos\vartheta_{\mathbf{p}\mathbf{q}}\right)\mathbf{J}_{\mathbf{p}}\cdot\mathbf{J}_{\mathbf{q}}  
\end{eqnarray} 

The evolution operator for the system represented by the Hamiltonian in Eq. (\ref{total})  
\begin{equation}
i\frac{\partial U}{\partial t} = \left( H_{\nu} + H_{\nu \nu} \right)  U ,
\end{equation}
can be approximately evaluated using the stationary phase approximation to its path integral 
representation \cite{Balantekin:2006tg}. This is equivalent to reducing the two-body Hamiltonian 
$H_{\nu \nu}$,  to a 
one-body one in an approximation similar to the random phase approximation (RPA). In this approximation the product of two commuting 
operators  $\hat{\cal O}_1$ and $\hat{\cal O}_2$ is approximated as 
\begin{equation}
\label{16a}
\hat{\cal O}_1 \hat{\cal O}_2 \sim 
\hat{\cal O}_1 \langle \hat{\cal O}_2 \rangle + \langle \hat{\cal O}_1 \rangle \hat{\cal O}_2 -
\langle \hat{\cal O}_1 \rangle \langle \hat{\cal O}_2 \rangle ,
\end{equation}
where the expectation values should be calculated with respect to a well-chosen state $|\Psi\rangle$ which satisfies 
the condition $ \langle \hat{\cal O}_1  \hat{\cal O}_2 \rangle = \langle \hat{\cal O}_1 \rangle \langle 
\hat{\cal O}_2\rangle~$. One then obtains the single-angle Hamiltonian 
\begin{equation}
\label{17}
\hat{H}\sim\hat{H}^{\mbox{\tiny RPA}} =  \sum_p\omega_{p}\hat{B}\cdot\mathbf{J}_p
+\vec{P}\cdot\mathbf{J}. 
\end{equation}
In writing Eq. (\ref{17}), assuming that the standard MSW effect is subdominant, matter terms are neglected. In this equation the polarization vector $\vec{P}$ was defined as
\begin{equation}
\label{18a}
\vec{P}_{\mathbf{p}}=2\langle\mathbf{J}_{\mathbf{p}}\rangle . 
\end{equation} 
If the operator averages in the above equations are calculated in the basis of the SU(2) coherent states associated with the flavor isospin,   
one obtains 
the standard reduced collective neutrino Hamiltonian, widely used in the 
literature \cite{Balantekin:2006tg}. 

In addition, both the full Hamiltonian and its one-body reduction possess an SU(N)$_f$ rotation 
symmetry in the N-neutrino flavor space \cite{Balantekin:2006tg,Duan:2008fd,Balantekin:2009dy}. 
It is not inconceivable that such a complex nonlinear system could exhibit further symmetries. A few of the conserved 
quantities in collective neutrino oscillations were already noted in the literature 
\cite{Raffelt:2007cb,Duan:2007mv}. To search for additional invariants, defining $\mu=\frac{\sqrt{2}G_{F}}{V}$, 
$\tau=\mu t$, and  $ \omega_{p}=\frac{1}{\mu}\frac{\delta m^{2}}{2p}$ \cite{Pehlivan:2010zz},  one can write the 
Hamiltonian of Eq. (\ref{total}) in the single-angle approximation as
\begin{equation}
\label{18}
\hat{H} = \sum_p\omega_{p}\hat{B}\cdot\mathbf{J}_p
+\mathbf{J}\cdot\mathbf{J} .
\end{equation}
Recall that in writing Eq. (\ref{18}), it was assumed that neutrino-neutrino interaction term is dominant and hence the matter 
terms were ignored. This Hamiltonian preserves the \emph{length of each spin}
\begin{equation}
\label{20}
\hat{L}_p=\mathbf{J}_{p}\cdot\mathbf{J}_{p}~,
\qquad\qquad \left[ \hat{H}, \hat{L}_p\right]=0~,
\end{equation}
as well as the \emph{total spin component} in the direction of the "external magnetic field", $\hat{B}$  
\begin{equation}
\label{21}
\hat{C}_0 =\hat{B}\cdot\mathbf{J}~, \qquad\qquad\quad \left[\hat{H},\hat{C}_0\right]=0~ . 
\end{equation}
The conservation law depicted in Eq. (\ref{20}) is not a new one, but the conservation of the total number of neutrinos 
with a given momentum: neither neutrino mixing nor coherent forward scattering of neutrinos off one another 
change the total number of neutrinos. In contrast, the conservation law depicted in Eq. (\ref{21}) is an additional one. 

The Hamiltonian in Eq. (\ref{18}) is similar to the reduced BCS Hamiltonian of the many-body theory. 
(However, one should note that the sign of the pairing term is opposite to that in the BCS Hamiltonian traditionally used in nuclear or condensed matter physics). 
One can exploit this duality to uncover the symmetries of the Hamiltonian. Already much work was done along this direction concerning the BCS theory. 
Eigenstates of the reduced BCS Hamiltonian were 
written by Richardson using a Bethe ansatz in \cite{Richardson1} and later generalized by 
Gaudin \cite{Gaudin1,Gaudin2}. Since the BCS Hamiltonian considered by Richardson is integrable, there should be 
constants of motion associated with it \cite{yuzb}. Using this analogy one can write down the constants of motion of the collective neutrino Hamiltonian in Eq. (\ref{total}) as \cite{Pehlivan:2011hp}
\begin{equation}
\label{suminvsum}
\hat{h}_{p} = \hat{B}\cdot\mathbf{J}_p+2\sum_{q\left(\neq p\right)}\frac{\mathbf{J}_{p}\cdot\mathbf{J}_{q}}{\omega_{p}-\omega_{q}}.
\end{equation}
The individual neutrino spin-lengths discussed above, $\hat{L}_p$,  are independent invariants. They are set by the initial 
conditions and are not changed by the evolution of the system under the collective Hamiltonian. 
On the other hand, the second invariant $\hat{C}_0$ is the sum of the invariants in Eq. (\ref{suminvsum}):
$
\hat{C}_0=\sum_{p}\hat{h}_{p}$. 
The Hamiltonian itself is also another linear combination of these invariants: 
\[
\hat{H}=\sum_{p}w_{p}\hat{h}_{p}+\sum_{p} \hat{L}_{p}~.
\]

Including antineutrinos, the conserved quantities for each neutrino energy mode $p$ take the form 
\begin{equation}
\hspace{-0.5cm}
\hat{h}_{p} = \hat{B}\cdot\mathbf{J}_p+2\sum_{q\left(\neq p\right)}\frac{\mathbf{J}_{p}\cdot\mathbf{J}_{q}}{\omega_{p}-\omega_{q}}+2\sum_{\bar{q}}\frac{\mathbf{J}_{p}\cdot\mathbf{\tilde{J}}_{\bar{q}}}{\omega_{p}-\omega_{\bar{q}}}, 
\end{equation}
where the quantities for antineutrinos have a tilde on them. Above we defined $ \omega_{\bar{p}}=-\frac{1}{\mu}\frac{\delta m^2}{2\bar{p}}$. 
Conserved quantities $\hat{h}_{\bar{p}}$ for different antineutrino energy modes are 
\begin{equation}
\hspace{-0.5cm}
\hat{h}_{\bar{p}} =\hat{B}\cdot\mathbf{\tilde{J}}_p+2\sum_{\bar{q}\left(\neq\bar{p}\right)}\frac{\mathbf{\tilde{J}}_{\bar{p}}\cdot\mathbf{\tilde{J}}_{\bar{q}}}{\omega_{\bar{p}}-\omega_{\bar{q}}}+2\sum_{q}\frac{\mathbf{\tilde{J}}_{\bar{p}}\cdot\mathbf{J}_{q}}{\omega_{\bar{p}}-\omega_{q}}~.
\end{equation}
The invariants of the reduced one-body Hamiltonian of Eq. (\ref{17}) can be written from those by taking the expectation value of the flavor isospin operators as 
\begin{equation}
\hspace{-0.5cm}
I_p=2\langle\hat{h}_{p}\rangle =\hat{B}\cdot\vec{P}_p+\sum_{q\left(\neq p\right)}\frac{\vec{P}_{p}\cdot\vec{P}_{q}}{\omega_{p}-\omega_{q}}+\sum_{\bar{q}}\frac{\vec{P}_{p}\cdot\vec{\tilde{P}}_{\bar{q}}}{\omega_{p}-\omega_{\bar{q}}}
\end{equation}
and 
\begin{equation}
\hspace{-0.5cm}
I_{\bar{p}}=2\langle\hat{h}_{\bar{p}}\rangle = \hat{B}\cdot\vec{\tilde{P}}_{\bar{p}}+\sum_{\bar{q}\left(\neq\bar{p}\right)}\frac{\vec{\tilde{P}}_{\bar{p}}\cdot\vec{\tilde{P}}_{\bar{q}}}{\omega_{\bar{p}}-\omega_{\bar{q}}}+\sum_{q}\frac{\vec{\tilde{P}}_{\bar{p}}\cdot\vec{P}_{q}}{\omega_{\bar{p}}-\omega_{q}} .
\end{equation}
It was shown that existence of such invariants could lead to collective neutrino oscillations \cite{Raffelt:2011yb}. 

As in the BCS theory, reduction to the one-body Hamiltonian of Eq. (\ref{17}) causes particle number conservation to fail. Particle 
number conservation can be reenforced by introducing a Lagrange multiplier: 
\begin{equation}
\hat{H}^{\mbox{\tiny RPA}} \rightarrow \hat{H}^{\mbox{\tiny RPA}}+\omega_c\mathbf{J}^0.  
\end{equation}
Diagonalization of this Hamiltonian gives rise to the phenomena called spectral split or swapping in the 
neutrino energy spectra \cite{Raffelt:2007cb,Duan:2007bt,Dasgupta:2009mg,Galais:2011gh} with 
the Lagrange multiplier playing the role of the swap frequency. 
To demonstrate this  we consider 
\begin{eqnarray}
\label{lm}
\hat{H}^{\mbox{\tiny RPA}}+\omega_c\mathbf{J}^0 &=& \sum_{p}(\omega_c-\omega_p)\mathbf{J}_p^0+\vec{{P}}\cdot\mathbf{J} \nonumber \\
&=& \sum_{\mathbf{p},s}  
2\lambda_p  \hat{U}^{\prime\dagger} \mathbf{J}^0_p \hat{U}^{\prime},  \nonumber
\end{eqnarray}
where the transforming operator is parameterized as 
\begin{equation}
\label{digtr}
\hat{U}^\prime=e^{\sum_p z_p J_p^+}\;e^{\sum_p \ln(1+|z_p|^2) J_p^0}\;e^{-\sum_p z_p^* J_p^-} 
\end{equation}
with 
\[
z_p=e^{i\delta}\tan{\theta_p}
\]
and 
\[
\cos{\theta_p} =\sqrt{\frac{1}{2}\left(1+\frac{\omega_c-\omega_p+ {P}^0}{2\lambda_p}\right)}. 
\]
This operator transforms matter-basis creation and annihilation operators into {\it quasi-particle} creation and annihilation operators:    
\begin{eqnarray}
&& \alpha_1(\mathbf{p},s) =\hat{U}^{\prime\dagger} a_1(\mathbf{p},s)\hat{U}^{\prime} \nonumber \\
&=& \cos{\theta_p} \; a_1(\mathbf{p},s) -e^{i\delta}\sin{\theta_p} \; a_2(\mathbf{p},s) \nonumber 
\end{eqnarray}
\begin{eqnarray}
&&\alpha_2(\mathbf{p},s)=\hat{U}^{\prime\dagger} a_2(\mathbf{p},s)\hat{U}^{\prime} \nonumber \\
&=&  e^{-i\delta}\sin{\theta_p} \; a_1(\mathbf{p},s)+\cos{\theta_p} \; a_2(\mathbf{p},s)\nonumber
\end{eqnarray}
so that we obtain a diagonal Hamiltonian: 
\begin{eqnarray}
&&\hat{H}^{\mbox{\tiny RPA}}+\omega_c\hat{J}^0 \nonumber \\ &=& \sum_{\mathbf{p},s}  
\lambda_p \left( \alpha_1^\dagger(\mathbf{p},s)\alpha_1(\mathbf{p},s)-\alpha_2^\dagger(\mathbf{p},s)\alpha_2(\mathbf{p},s)   \right) . 
\nonumber
\end{eqnarray}

Let us assume that initially ($\lim \mu \rightarrow \infty$) there are more $\nu_e$'s and all neutrinos are in flavor eigenstates. We then have 
\[
\lim \cos\theta_p = \lim \sqrt{\frac{1}{2}\left(1+\frac{P^0}{|\vec{P}|}\cos{2\theta}\right)} = \cos \theta,
\]
i.e., the diagonalizing transformation of Eq. (\ref{digtr}) reduces into the neutrino mixing transformation of Eq. (\ref{mixalg}) and the total Hamiltonian of Eq. (\ref{lm}) is diagonalized by the flavor eigenstates:  
\[
\alpha_1(\mathbf{p},s)=\hat{U}^\dagger a_1(\mathbf{p},s)\hat{U} = a_e(\mathbf{p},s) .
\]
After neutrinos propagate to a region with very low neutrino density ($\mu \rightarrow 0$) one gets  
\[
\cos\theta_p=\sqrt{\frac{1}{2}\left(1+\frac{\omega_c-\omega_p}{|\omega_c-\omega_p|}\right)} 
\Rightarrow \left\{ \begin{array}{rl} 
1 &  \omega_p < \omega_c\\
 0 & \omega_p > \omega_c
\end{array} \right.
\]
yielding
\[
\alpha_1(\mathbf{p},s)=\hat{U}^\dagger a_1(\mathbf{p},s)\hat{U} \Rightarrow \left\{ \begin{array}{rl} 
a_1(\mathbf{p},s)  &  \omega_p < \omega_c\\
- a_2(\mathbf{p},s)  & \omega_p > \omega_c
\end{array} \right. ,
\]
i.e. neutrinos with $\omega_p < \omega_c$ and $\omega_p > \omega_c$ evolve into different mass eigenstates. In Ref. \cite{Raffelt:2007cb} it was shown that such an evolution leads to a phenomenon called spectral split or spectral swapping.  
  
The fact that invariants of the full Hamiltonian are also invariants of the one-body Hamiltonian of Eq. (\ref{17}) 
when they are properly linearized provides confidence in the aptness of the linearization procedure itself. 
One should also note that 
a rather different linearization procedure has been used to carry out flavor-stability analysis of dense neutrino streams 
\cite{Banerjee:2011fj}. 

Recent numerical work with three flavors in the multi-angle approximation uncovers significant differences between single- and multi-angle formulations \cite{Cherry:2010yc}. In particular, multi-angle formulation is found to reduce the adiabaticity of flavor evolution in the normal neutrino mass hierarchy, resulting in lower swap energies. Thus it seems that single-angle approximation seems to be sufficient in some cases, but inadequate in other situations.  

\section{Conclusions}

Despite much work by many authors still a number of questions remain regarding the many-body behavior of the neutrinos in core-collapse supernovae. For example, in the calculations so far neutrinos are assumed to be emitted half-isotropically (only outward-moving modes are occupied with backward-moving modes being empty). However, recent realistic supernova simulations suggest that neutrino angular distributions are not half-isotropic \cite{Ott:2008jb}. Flavor-dependent angular distributions may lead to multi-angle instabilities 
\cite{Sawyer:2008zs,Mirizzi:2011tu}.  
Neutrinos that scatter in non-forward directions could create a "neutrino halo" that would interact with the other outgoing neutrinos. 
It was also argued that fraction of outflowing neutrinos interacting with this neutrino halo is significant  \cite{Cherry:2012zw} . 

It would be desirable to verify the assumption that contributions beyond the ÒRPAÓ-mean field are indeed small by explicitly calculating them. One can ask if there are other roles symmetries play in this collective system. This is not only of academic interest but also a practical one 
as judicious use of the symmetries would help numerical calculations of this complex system. We should further look for other observable signatures of the neutrino-neutrino terms besides spectral splits and swappings. 
Future work should try to answer some of these questions.

\section*{Acknowledgments}
I would like to thank J. M. Fetter, G. M. Fuller, J. Gava, T. Kajino, G. C. McLaughlin, F. N. Loreti, T. Maruyama, Y. Pehlivan, Y. Qian, C. Volpe, T. Yoshida, and H. Yuksel. Much of the work presented here was carried out in collaboration with them.  
This work was supported in part
by the U.S. National Science Foundation Grants No. 
PHY-0855082 and PHY-1205024.
and
in part by the University of Wisconsin Research Committee with funds
granted by the Wisconsin Alumni Research Foundation.

%% \-------------------------------------------------------------------

%% References

%% --- EDIT and comment out to use here if you want to use BibTeX ---
%\nocite{*}
%\bibliographystyle{elsarticle-num}
%\bibliography{martin}

\begin{thebibliography}{00}

%\cite{Duan:2009cd}
\bibitem{Duan:2009cd} 
  H.~Duan and J.~P. Kneller,
  %``Neutrino flavour transformation in supernovae,''
  J.\ Phys.\ G {\bf 36}, 113201 (2009)
  [arXiv:0904.0974 [astro-ph.HE]].
  %%CITATION = ARXIV:0904.0974;%%

%\cite{Raffelt:2010zza}
\bibitem{Raffelt:2010zza} 
  G.~G.~Raffelt,
  %``Physics opportunities with supernova neutrinos,''
  Prog.\ Part.\ Nucl.\ Phys.\  {\bf 64}, 393 (2010).
  %%CITATION = PPNPD,64,393;%%

%\cite{Duan:2010bg}
\bibitem{Duan:2010bg} 
  H.~Duan, G.~M.~Fuller and Y.~-Z.~Qian,
  %``Collective Neutrino Oscillations,''
  Ann.\ Rev.\ Nucl.\ Part.\ Sci.\  {\bf 60}, 569 (2010)
  [arXiv:1001.2799 [hep-ph]].
  %%CITATION = ARXIV:1001.2799;%%

%\cite{Balantekin:2003ip}
\bibitem{Balantekin:2003ip} 
  A.~B.~Balantekin and G.~M.~Fuller,
  %``Supernova neutrino - nucleus astrophysics,''
  J.\ Phys.\ G  {\bf 29}, 2513 (2003)
  [astro-ph/0309519].
  %%CITATION = ASTRO-PH/0309519;%% 

%\cite{McLaughlin:1997qi}
\bibitem{McLaughlin:1997qi} 
  G.~C.~McLaughlin, G.~M.~Fuller and J.~R.~Wilson,
  %``The Influence of nuclear composition on the electron fraction in the post-core-bounce supernova environment,''
  Astrophys.\ J.\  {\bf 472}, 440 (1996)
  [astro-ph/9701114]; 
  %%CITATION = ASTRO-PH/9701114;%%
%\cite{Meyer:1998sn}
%\bibitem{Meyer:1998sn} 
  B.~S.~Meyer, G.~C.~McLaughlin and G.~M.~Fuller,
  %``Neutrino capture and r process nucleosynthesis,''
  Phys.\ Rev.\ C {\bf 58}, 3696 (1998)
  [astro-ph/9809242].
  %%CITATION = ASTRO-PH/9809242;%%
  
%\cite{Wolfenstein:1977ue}
\bibitem{Wolfenstein:1977ue} 
  L.~Wolfenstein,
  %``Neutrino Oscillations in Matter,''
  Phys.\ Rev.\ D {\bf 17}, 2369 (1978).
  %%CITATION = PHRVA,D17,2369;%%  
  
%\cite{Botella:1986wy}
\bibitem{Botella:1986wy}
  F.~J.~Botella, C.~S.~Lim and W.~J.~Marciano,
  %``RADIATIVE CORRECTIONS TO NEUTRINO INDICES OF REFRACTION,''
  Phys.\ Rev.\  D {\bf 35}, 896 (1987).
  %%CITATION = PHRVA,D35,896;%%

%\cite{Akhmedov:2002zj}
\bibitem{Akhmedov:2002zj} 
  E.~K.~Akhmedov, C.~Lunardini and A.~Y.~.Smirnov,
  %``Supernova neutrinos: Difference of muon-neutrino - tau-neutrino fluxes and conversion effects,''
  Nucl.\ Phys.\ B {\bf 643}, 339 (2002)
  [hep-ph/0204091].
  %%CITATION = HEP-PH/0204091;%
  
%\cite{Loreti:1994ry}
\bibitem{Loreti:1994ry} 
  F.~N.~Loreti and A.~B.~Balantekin,
  %``Neutrino oscillations in noisy media,''
  Phys.\ Rev.\ D {\bf 50}, 4762 (1994)
  [nucl-th/9406003]; 
  %%CITATION = NUCL-TH/9406003;%%
%\cite{Balantekin:1996pp}
%\bibitem{Balantekin:1996pp} 
  A.~B.~Balantekin, J.~M.~Fetter and F.~N.~Loreti,
  %``The MSW effect in a fluctuating matter density,''
  Phys.\ Rev.\ D {\bf 54}, 3941 (1996)
  [astro-ph/9604061].
  %%CITATION = ASTRO-PH/9604061;%%
  
  %\cite{Loreti:1995ae}
\bibitem{Loreti:1995ae} 
  F.~N.~Loreti, Y.~Z.~Qian, G.~M.~Fuller and A.~B.~Balantekin,
  %``Effects of random density fluctuations on matter enhanced neutrino flavor transitions in supernovae and implications for supernova %dynamics and nucleosynthesis,''
  Phys.\ Rev.\ D {\bf 52}, 6664 (1995)
  [astro-ph/9508106].
  %%CITATION = ASTRO-PH/9508106;%%

%\cite{Cherry:2011fm}
\bibitem{Cherry:2011fm} 
  J.~F.~Cherry, M.~-R.~Wu, J.~Carlson, H.~Duan, G.~M.~Fuller and Y.~-Z.~Qian,
  %``Density Fluctuation Effects on Collective Neutrino Oscillations in O-Ne-Mg Core-Collapse Supernovae,''
  Phys.\ Rev.\ D {\bf 84}, 105034 (2011)
  [arXiv:1108.4064 [astro-ph.HE]]; 
  %%CITATION = ARXIV:1108.4064;%%
%\cite{Reid:2011zz}
%\bibitem{Reid:2011zz} 
  G.~Reid, J.~Adams and S.~Seunarine,
  %``Collective neutrino oscillations in turbulent backgrounds,''
  Phys.\ Rev.\ D {\bf 84}, 085023 (2011); 
  %%CITATION = PHRVA,D84,085023;%%
%\cite{Kneller:2010sc}
%\bibitem{Kneller:2010sc} 
  J.~P.~Kneller and C.~Volpe,
  %``Turbulence effects on supernova neutrinos,''
  Phys.\ Rev.\ D {\bf 82}, 123004 (2010)
  [arXiv:1006.0913 [hep-ph]]; 
  %%CITATION = ARXIV:1006.0913;%%
%\cite{Friedland:2006ta}
%\bibitem{Friedland:2006ta} 
  A.~Friedland and A.~Gruzinov,
  %``Neutrino signatures of supernova turbulence,''
  astro-ph/0607244.
  %%CITATION = ASTRO-PH/0607244;%%

%\cite{Balantekin:2007es}
\bibitem{Balantekin:2007es}
  A.B. Balantekin, J. Gava and C. Volpe,  
  %``Possible CP-Violation effects in core-collapse Supernovae,''
  {\it Phys.\ Lett.\  B} {\bf 662}, 396 (2008)
  (Preprint arXiv:0710.3112 [astro-ph]); 
  %%CITATION = PHLTA,B662,396;%%

%\cite{Kneller:2009vd}
\bibitem{Kneller:2009vd}
  J.P. Kneller and G.C. McLaughlin  
  %``Three Flavor Neutrino Oscillations in Matter: Flavor Diagonal Potentials,
  %the Adiabatic Basis and the CP phase,''
  {\it Phys.\ Rev.\  D} {\bf 80}, 053002 (2009)
  (Preprint arXiv:0904.3823 [hep-ph]).
  %%CITATION = PHRVA,D80,053002;%%

%\cite{Palazzo:2011rj}
\bibitem{Palazzo:2011rj}
  A. Palazzo 
  %``Testing the very-short-baseline neutrino anomalies at the solar sector,''
  {\it Phys.\ Rev.\  D} {\bf 83}, 113013 (2011)
  [arXiv:1105.1705 [hep-ph]].
  %%CITATION = PHRVA,D83,113013;%%

%\cite{Gava:2008rp}
\bibitem{Gava:2008rp} 
  J.~Gava and C.~Volpe,
  %``Collective neutrinos oscillation in matter and CP-violation,''
  Phys.\ Rev.\ D {\bf 78}, 083007 (2008)
  [arXiv:0807.3418 [astro-ph]].
  %%CITATION = ARXIV:0807.3418;%%

%\cite{Mention:2011rk}
\bibitem{Mention:2011rk} 
  G.~Mention, M.~Fechner, T.~.Lasserre, T.~.A.~Mueller, D.~Lhuillier, M.~Cribier and A.~Letourneau,
  %``The Reactor Antineutrino Anomaly,''
  Phys.\ Rev.\ D {\bf 83}, 073006 (2011)
  [arXiv:1101.2755 [hep-ex]].
  %%CITATION = ARXIV:1101.2755;%%
  
%\cite{Abazajian:2012ys}
\bibitem{Abazajian:2012ys} 
  K.~N.~Abazajian, M.~A.~Acero, S.~K.~Agarwalla, A.~A.~Aguilar-Arevalo, C.~H.~Albright, S.~Antusch, C.~A.~Arguelles and A.~B.~Balantekin {\it et al.},
  %``Light Sterile Neutrinos: A White Paper,''
  arXiv:1204.5379 [hep-ph].
  %%CITATION = ARXIV:1204.5379;%%

%\cite{Giunti:2011cp}
\bibitem{Giunti:2011cp} 
  C.~Giunti and M.~Laveder,
  %``Implications of 3+1 Short-Baseline Neutrino Oscillations,''
  Phys.\ Lett.\ B {\bf 706}, 200 (2011)
  [arXiv:1111.1069 [hep-ph]].
  %%CITATION = ARXIV:1111.1069;%%  
  
  %\cite{McLaughlin:1999pd}
\bibitem{McLaughlin:1999pd} 
  G.~C.~McLaughlin, J.~M.~Fetter, A.~B.~Balantekin and G.~M.~Fuller,
  %``An Active sterile neutrino transformation solution for r process nucleosynthesis,''
  Phys.\ Rev.\ C {\bf 59}, 2873 (1999)
  [astro-ph/9902106]; 
  %%CITATION = ASTRO-PH/9902106;%%
%\cite{Caldwell:1999zk}
%\bibitem{Caldwell:1999zk} 
  D.~O.~Caldwell, G.~M.~Fuller and Y.~-Z.~Qian,
  %``Sterile neutrinos and supernova nucleosynthesis,''
  Phys.\ Rev.\ D {\bf 61}, 123005 (2000)
  [astro-ph/9910175]; 
  %%CITATION = ASTRO-PH/9910175;%%
  %\cite{Fetter:2002xx}
%\bibitem{Fetter:2002xx} 
  J.~Fetter, G.~C.~McLaughlin, A.~B.~Balantekin and G.~M.~Fuller,
  %``Active sterile neutrino conversion: Consequences for the r process and supernova neutrino detection,''
  Astropart.\ Phys.\  {\bf 18}, 433 (2003)
  [hep-ph/0205029].
  %%CITATION = HEP-PH/0205029;%%

%\cite{Tamborra:2011is}
\bibitem{Tamborra:2011is} 
  I.~Tamborra, G.~G.~Raffelt, L.~Hudepohl and H.~-T.~Janka,
  %``Impact of eV-mass sterile neutrinos on neutrino-driven supernova outflows,''
  JCAP {\bf 1201}, 013 (2012)
  [arXiv:1110.2104 [astro-ph.SR]].
  %%CITATION = ARXIV:1110.2104;%%

%\cite{Hidaka:2007se}
\bibitem{Hidaka:2007se} 
  J.~Hidaka and G.~M.~Fuller,
  %``Sterile Neutrino-Enhanced Supernova Explosions,''
  Phys.\ Rev.\ D {\bf 76}, 083516 (2007)
  [arXiv:0706.3886 [astro-ph]].
  %%CITATION = ARXIV:0706.3886;%%

%\cite{Beacom:2010kk}
\bibitem{Beacom:2010kk} 
  J.~F.~Beacom,
  %``The Diffuse Supernova Neutrino Background,''
  Ann.\ Rev.\ Nucl.\ Part.\ Sci.\  {\bf 60}, 439 (2010)
  [arXiv:1004.3311 [astro-ph.HE]].
  %%CITATION = ARXIV:1004.3311;%%

%\cite{Balantekin:2006tg}
\bibitem{Balantekin:2006tg} 
  A.~B.~Balantekin and Y.~Pehlivan,
  %``Neutrino-Neutrino Interactions and Flavor Mixing in Dense Matter,''
  J.\ Phys.\ G {\bf 34}, 47 (2007)
  [astro-ph/0607527].
  %%CITATION = ASTRO-PH/0607527;%%

\bibitem{Pantaleone:1992eq}
  J.~T.~Pantaleone,
  %``Neutrino oscillations at high densities,''
  Phys.\ Lett.\ B {\bf 287}, 128 (1992); 
  %%CITATION = PHLTA,B287,128;%%
%\bibitem{Pantaleone:1992xh}
%  J.~T.~Pantaleone,
%   ``Dirac neutrinos in dense matter,''
  Phys.\ Rev.\ D {\bf 46}, 510 (1992).
  %%CITATION = PHRVA,D46,510;%%

\bibitem{Qian:1994wh}
  Y.~Z.~Qian and G.~M.~Fuller,
  %``Neutrino-neutrino scattering and matter enhanced neutrino flavor
  %transformation in Supernovae,''
  Phys.\ Rev.\ D {\bf 51}, 1479 (1995)
  [arXiv:astro-ph/9406073].
  %%CITATION = ASTRO-PH 9406073;%%

\bibitem{Pastor:2002we}
  S.~Pastor and G.~Raffelt,
  %``Flavor oscillations in the supernova hot bubble region: Nonlinear  effects
  %of neutrino background,''
  Phys.\ Rev.\ Lett.\  {\bf 89}, 191101 (2002)
  [arXiv:astro-ph/0207281].
  %%CITATION = ASTRO-PH 0207281;%%

\bibitem{Friedland:2003dv}
  A.~Friedland and C.~Lunardini,
%   ``Neutrino flavor conversion in a neutrino background: Single- versus
%   multi-particle description,''
  Phys.\ Rev.\ D {\bf 68}, 013007 (2003)
  [arXiv:hep-ph/0304055].
  %%CITATION = HEP-PH 0304055;%%

\bibitem{Balantekin:2004ug}
  A.~B.~Balantekin and H.~Yuksel,
  %``Neutrino mixing and nucleosynthesis in core-collapse supernovae,''
  New J.\ Phys.\  {\bf 7}, 51 (2005)
  [arXiv:astro-ph/0411159].
  %%CITATION = ASTRO-PH 0411159;%%

\bibitem{Duan:2005cp}
  H.~Duan, G.~M.~Fuller and Y.~Z.~Qian,
  %``Collective neutrino flavor transformation in supernovae,''
  Phys.\ Rev.\  D {\bf 74}, 123004 (2006)
  [arXiv:astro-ph/0511275]
  %%CITATION = PHRVA,D74,123004;%%

%\cite{Hannestad:2006nj}
\bibitem{Hannestad:2006nj}
  S.~Hannestad, G.~G.~Raffelt, G.~Sigl and Y.~Y.~Y.~Wong,
  %``Self-induced conversion in dense neutrino gases: Pendulum in flavour
  %space,''
  Phys.\ Rev.\  D {\bf 74}, 105010 (2006)
  [Erratum-ibid.\  D {\bf 76}, 029901 (2007)]
  [arXiv:astro-ph/0608695].
  %%CITATION = PHRVA,D74,105010;%%
  
%\cite{Raffelt:2007cb}
\bibitem{Raffelt:2007cb}
  G.~G.~Raffelt and A.~Y.~Smirnov,
  %``Self-induced spectral splits in supernova neutrino fluxes,''
  Phys.\ Rev.\  D {\bf 76}, 081301 (2007)
  [Erratum-ibid.\  D {\bf 77}, 029903 (2008)]
  [arXiv:0705.1830 [hep-ph]]; 
  %%CITATION = PHRVA,D76,081301;%%
 %``Adiabaticity and spectral splits in collective neutrino transformations,'' 
  Phys.\ Rev.\  D {\bf 76}, 125008 (2007) 
  [arXiv:0709.4641 [hep-ph]]. 
  %%CITATION = PHRVA,D76,125008;%% 

\bibitem{Friedland:2006ke}
  A.~Friedland, B.~H.~J.~McKellar and I.~Okuniewicz,
%   ``Construction and analysis of a simplified many-body neutrino model,''
  %
  Phys.\ Rev.\ D {\bf 73}, 093002 (2006)
  [arXiv:hep-ph/0602016].
  %%CITATION = HEP-PH 0602016;%%
  
%\cite{Duan:2007mv}
\bibitem{Duan:2007mv} 
  H.~Duan, G.~M.~Fuller, J.~Carlson and Y.~-Z.~Qian,
  %``Analysis of Collective Neutrino Flavor Transformation in Supernovae,''
  Phys.\ Rev.\ D {\bf 75}, 125005 (2007)
  [astro-ph/0703776].
  %%CITATION = ASTRO-PH/0703776;%%
  
 %\cite{Fogli:2007bk}
\bibitem{Fogli:2007bk} 
  G.~L.~Fogli, E.~Lisi, A.~Marrone and A.~Mirizzi,
  %``Collective neutrino flavor transitions in supernovae and the role of trajectory averaging,''
  JCAP {\bf 0712}, 010 (2007)
  [arXiv:0707.1998 [hep-ph]].
  %%CITATION = ARXIV:0707.1998;%%

%\cite{Chakraborty:2009ej}
\bibitem{Chakraborty:2009ej} 
  S.~Chakraborty, S.~Choubey, S.~Goswami and K.~Kar,
  %``Collective Flavor Oscillations Of Supernova Neutrinos and r-Process Nucleosynthesis,''
  JCAP {\bf 1006}, 007 (2010)
  [arXiv:0911.1218 [hep-ph]].
  %%CITATION = ARXIV:0911.1218;%%
  
%\cite{Dasgupta:2011jf}
\bibitem{Dasgupta:2011jf} 
  B.~Dasgupta, E.~P.~O'Connor and C.~D.~Ott,
  %``The Role of Collective Neutrino Flavor Oscillations in Core-Collapse Supernova Shock Revival,''
  Phys.\ Rev.\ D {\bf 85}, 065008 (2012)
  [arXiv:1106.1167 [astro-ph.SR]].
  %%CITATION = ARXIV:1106.1167;%%    

%\cite{Mirizzi:2011tu}
\bibitem{Mirizzi:2011tu} 
  A.~Mirizzi and P.~D.~Serpico,
  %``Instability in the dense supernova neutrino gas with flavor-dependent angular distributions,''
  Phys.\ Rev.\ Lett.\  {\bf 108}, 231102 (2012)
  [arXiv:1110.0022 [hep-ph]].
  %%CITATION = ARXIV:1110.0022;%%
  
%\cite{Baldo:2012hp}
\bibitem{Baldo:2012hp} 
  M.~Baldo and V.~Palmisano,
  %``Single-angle to multi-angle transition in the collective flavor dynamics of neutrinos in supernovae,''
  arXiv:1202.2243 [astro-ph.HE].
  %%CITATION = ARXIV:1202.2243;%%  
  
%\cite{Saviano:2012yh}
\bibitem{Saviano:2012yh} 
  N.~Saviano, S.~Chakraborty, T.~Fischer and A.~Mirizzi,
  %``Stability analysis of collective neutrino oscillations in the supernova accretion phase with realistic energy and angle distributions,''
  Phys.\ Rev.\ D {\bf 85}, 113002 (2012)
  [arXiv:1203.1484 [hep-ph]].
  %%CITATION = ARXIV:1203.1484;%%  

%\cite{Balantekin:2011ft}
\bibitem{Balantekin:2011ft} 
  A.~B.~Balantekin and A.~Malkus,
  %``Solar Neutrino Matter Effects Redux,''
  Phys.\ Rev.\ D {\bf 85}, 013010 (2012)
  [arXiv:1109.5216 [hep-ph]].
  %%CITATION = ARXIV:1109.5216;%%

%\cite{Kostelecky:1993ys}
\bibitem{Kostelecky:1993ys}
  V.~A.~Kostelecky and S.~Samuel,
  %``Nonlinear Neutrino Oscillations In The Expanding Universe,''
  Phys.\ Rev.\  D {\bf 49}, 1740 (1994).
  %%CITATION = PHRVA,D49,1740;%%

\bibitem{Abazajian:2002qx}
  K.~N.~Abazajian, J.~F.~Beacom and N.~F.~Bell,
%   ``Stringent constraints on cosmological neutrino antineutrino asymmetries
%   from synchronized flavor transformation,''
  %
  Phys.\ Rev.\ D {\bf 66}, 013008 (2002)
  [arXiv:astro-ph/0203442].
  %%CITATION = ASTRO-PH 0203442;%%

%\cite{Sawyer:2005jk}
\bibitem{Sawyer:2005jk} 
  R.~F.~Sawyer,
  %``Speed-up of neutrino transformations in a supernova environment,''
  Phys.\ Rev.\ D {\bf 72}, 045003 (2005)
  [hep-ph/0503013].
  %%CITATION = HEP-PH/0503013;%%

%\cite{Duan:2008fd}
\bibitem{Duan:2008fd}
  H.~Duan, G.~M.~Fuller and Y.~Z.~Qian,
  %``Symmetries in collective neutrino oscillations,''
  J.\ Phys.\ G {\bf 36}, 105003 (2009)
  [arXiv:0808.2046 [astro-ph]].

%\cite{Balantekin:2009dy}
\bibitem{Balantekin:2009dy} 
  A.~B.~Balantekin,
  %``Neutrinos and Symmetries,''
  Nucl.\ Phys.\ A {\bf 844}, 14C (2010)
  [arXiv:0910.1814 [nucl-th]].
  %%CITATION = ARXIV:0910.1814;%%

%\cite{Pehlivan:2010zz}
\bibitem{Pehlivan:2010zz} 
  Y.~Pehlivan, T.~Kajino, A.~B.~Balantekin, T.~Yoshida and T.~Maruyama,
  %``On the neutrino self refraction problem from a many-body perspective,''
  \emph{AIP Conf.\ Proc.}  \textbf{1269}, 189-194 (2010).
  %%CITATION = APCPC,1269,189;%%

\bibitem{Richardson1}
  R.W. Richardson,   Phys. Lett. \textbf{3} (1963) 277.
  
 \bibitem{Gaudin1}
  M. Gaudin, 
  J. Physique \textbf{37}(1976), 1087.

\bibitem{Gaudin2}
  M. Gaudin, 
  {\it La Fonction d'onde de Bethe},  
  Collection du Commissariat a l'\'{e}nergie atomique, Masson, Paris, 1983.

\bibitem{yuzb} 
 A.A. Yuzbashyan, B.L. Altshuler, V.B. Kuznetsov, and V.E. Enolskii, J. Phys. A: Math. Gen. {\bf 38}, 7831 (2005).   

%\cite{Pehlivan:2011hp}
\bibitem{Pehlivan:2011hp} 
  Y.~Pehlivan, A.~B.~Balantekin, T.~Kajino and T.~Yoshida,
  %``Invariants of Collective Neutrino Oscillations,''
  Phys.\ Rev.\ D {\bf 84}, 065008 (2011)
  [arXiv:1105.1182 [astro-ph.CO]].
  %%CITATION = ARXIV:1105.1182;%%

%\cite{Raffelt:2011yb}
\bibitem{Raffelt:2011yb} 
  G.~G.~Raffelt,
  %``N-mode coherence in collective neutrino oscillations,''
  Phys.\ Rev.\ D {\bf 83}, 105022 (2011)
  [arXiv:1103.2891 [hep-ph]].
  %%CITATION = ARXIV:1103.2891;%%

%\cite{Duan:2007bt}
\bibitem{Duan:2007bt} 
  H.~Duan, G.~M.~Fuller, J.~Carlson and Y.~-Q.~Zhong,
  %``Neutrino Mass Hierarchy and Stepwise Spectral Swapping of Supernova Neutrino Flavors,''
  Phys.\ Rev.\ Lett.\  {\bf 99}, 241802 (2007)
  [arXiv:0707.0290 [astro-ph]].
  %%CITATION = ARXIV:0707.0290;%%

%\cite{Dasgupta:2009mg}
\bibitem{Dasgupta:2009mg} 
  B.~Dasgupta, A.~Dighe, G.~G.~Raffelt and A.~Y.~.Smirnov,
  %``Multiple Spectral Splits of Supernova Neutrinos,''
  Phys.\ Rev.\ Lett.\  {\bf 103}, 051105 (2009)
  [arXiv:0904.3542 [hep-ph]].
  %%CITATION = ARXIV:0904.3542;%%

%\cite{Galais:2011gh}
\bibitem{Galais:2011gh} 
  S.~Galais and C.~Volpe,
  %``The neutrino spectral split in core-collapse supernovae: a magnetic resonance phenomenon,''
  Phys.\ Rev.\ D {\bf 84}, 085005 (2011)
  [arXiv:1103.5302 [astro-ph.SR]].
  %%CITATION = ARXIV:1103.5302;%%

%\cite{Banerjee:2011fj}
\bibitem{Banerjee:2011fj} 
  A.~Banerjee, A.~Dighe and G.~Raffelt,
  %``Linearized flavor-stability analysis of dense neutrino streams,''
  Phys.\ Rev.\ D {\bf 84}, 053013 (2011)
  [arXiv:1107.2308 [hep-ph]].
  %%CITATION = ARXIV:1107.2308;%%

%\cite{Cherry:2010yc}
\bibitem{Cherry:2010yc}
  J.~F.~Cherry, G.~M.~Fuller, J.~Carlson, H.~Duan, Y.~-Z.~Qian,
  %``Multi-Angle Simulation of Flavor Evolution in the Neutrino Neutronization Burst From an O-Ne-Mg Core-Collapse Supernova,''
  Phys.\ Rev.\  {\bf D82}, 085025 (2010).
  [arXiv:1006.2175 [astro-ph.HE]].

%\cite{Ott:2008jb}
\bibitem{Ott:2008jb}
  C.~D.~Ott, A.~Burrows, L.~Dessart and E.~Livne,
  %``2D Multi-Angle, Multi-Group Neutrino Radiation-Hydrodynamic Simulations of
  %Postbounce Supernova Cores,''
  Astrophys.\ J.\  {\bf 685}, 1069 (2008)
  [arXiv:0804.0239 [astro-ph]]; 
  %%CITATION = ASJOA,685,1069;%%
%\cite{Chakraborty:2011gd}
%\bibitem{Chakraborty:2011gd}
  S.~Chakraborty, T.~Fischer, A.~Mirizzi, N.~Saviano and R.~Tomas,
  %``Analysis of matter suppression in collective neutrino oscillations during
  %the supernova accretion phase,''
  Phys.\ Rev.\  D {\bf 84}, 025002 (2011)
  [arXiv:1105.1130 [hep-ph]].
  %%CITATION = PHRVA,D84,025002;%%

%\cite{Sawyer:2008zs}
\bibitem{Sawyer:2008zs}
  R.~F.~Sawyer,
  %``The multi-angle instability in dense neutrino systems,''
  Phys.\ Rev.\  {\bf D79}, 105003 (2009).
  [arXiv:0803.4319 [astro-ph]].

%\cite{Cherry:2012zw}
\bibitem{Cherry:2012zw} 
  J.~F.~Cherry, J.~Carlson, A.~Friedland, G.~M.~Fuller and A.~Vlasenko,
  %``Neutrino scattering and flavor transformation in supernovae,''
  Phys.\ Rev.\ Lett.\  {\bf 108}, 261104 (2012)
  [arXiv:1203.1607 [hep-ph]].
  %%CITATION = ARXIV:1203.1607;%%

\end{thebibliography}


%% --- EDIT Bibriography ---

%% --------------------------------

\end{document}